\documentclass[10pt,conference]{IEEEtran}
\IEEEoverridecommandlockouts
\usepackage{cite}
\usepackage{subcaption}
\usepackage{amsmath,amssymb,amsfonts}
\usepackage{algorithm}
\usepackage{algpseudocode}
\usepackage{url} 
\usepackage{hyperref} 
\usepackage{graphicx}
\usepackage{textcomp}
\usepackage{xcolor}
\usepackage{algorithm}
\usepackage{booktabs}
\usepackage{microtype}
\usepackage{listings}
\usepackage{multirow}
\usepackage[most, breakable]{tcolorbox}
\usepackage{flushend}
\usepackage{todonotes}

\newcommand{\lt}{\ensuremath{<}} % Define \lt to represent the less-than symbol in math mode
\newcommand{\gt}{\ensuremath{>}} % Define \gt for the greater-than symbol as well

\def\BibTeX{{\rm B\kern-.05em{\sc i\kern-.025em b}\kern-.08em
    T\kern-.1667em\lower.7ex\hbox{E}\kern-.125emX}}
\begin{document}

\title{Can AI Agents Generate Microservices? \\How Far are We?
}

% \and
% \and
% \and
% \and
% \and

\author{Bassam Adnan$^{1}$, Matteo Esposito$^2$, Davide Taibi$^{2,3}$, Karthik Vaidhyanathan$^1$,  \\$^1$\textit{SERC, IIIT Hyderabad}, India --- $^2$\textit{University of Oulu}, Finland --- $^3$\textit{University of Southern Denmark}, Denmark\\ bassam.adnan@research.iiit.ac.in; karthik.vaidhyanathan@iiit.ac.in; matteo.esposito@oulu.fi; taibi@imada.sdu.dk }

% \author{Anonymous Author(s)}

\maketitle

\begin{abstract}\textit{Context.} LLMs have advanced code generation, but their use for generating microservices with explicit dependencies and API contracts remains understudied.

\textit{Goal.} We examine whether AI agents can generate functional microservices and how different forms of contextual information influence their performance.

\textit{Method.} We assess 144 generated microservices across 3 agents, 4 projects, 2 prompting strategies, and 2 scenarios. Incremental generation operates within existing systems and is evaluated with unit tests. Clean state generation starts from requirements alone and is evaluated with integration tests. We analyze functional correctness, code quality, and efficiency.

\textit{Results.} Minimal prompts outperformed detailed ones in incremental generation, with 50–76\% unit test pass rates. Clean state generation produced higher integration test pass rates (81–98\%), indicating strong API contract adherence. Generated code showed lower complexity than human baselines. Generation times varied widely across agents, averaging 6–16 minutes per service.

\textit{Conclusions.} AI agents can produce microservices with maintainable code, yet inconsistent correctness and reliance on human oversight show that fully autonomous microservice generation is not yet achievable.

\end{abstract}


\section{Introduction}

Automating the generation of executable systems from architecture descriptions has long been a central ambition in Software Architecture (SA), offering benefits such as stronger compliance and improved traceability~\cite{garlan1993_component_connector}. The SA community has explored this vision through Architecture Description Languages (ADLs) and Domain-Specific Languages (DSLs)\cite{malavolta2012industry}, yet adoption has remained limited due to steep learning curves and insufficient tool support\cite{hutchinson2011mde}.

Recent advances in AI offer a new direction. Large Language Models (LLMs) now perform a wide range of Software Engineering (SE) tasks~\cite{hou2024llm4se}, including code completion, function generation, and program repair~\cite{chen2021codex,svyatkovskiy2020intellicode,nijkamp2023codegen,esposito_large_2024}. Research on GenAI for software architecture has begun to explore requirements-to-architecture and architecture-to-code transitions, yet only few studies attempt the full requirements-to-architecture-to-code pipeline, and most lack robust validation strategies for AI-generated artifacts~\cite{esposito2025genai}. Work on generating larger architectural units remains scarce, with early efforts emerging in specific contexts such as serverless functions~\cite{arun}.

A broader shift is now occurring from standalone LLMs to AI agents. These agents augment foundation models with tool usage, memory, and iterative refinement workflows~\cite{wang2024survey,yang2024sweagent}. Unlike single-pass LLMs, agents can run tests, analyze compilation errors, use repository context, and iteratively refine their outputs, making them more suitable for automatic code generation.~\cite{xia2024agent,swebench,bouzenia2025understandingsoftwareengineeringagents}. Tools such as Codex~\cite{chen2021codex} and Cursor~\cite{cursorAI} are already being integrated into real development environments. However, deployment has outpaced empirical understanding: incidents like Replit’s agent deleting a production database\footnote{\url{https://x.com/amasad/status/1946986468586721478}} illustrate the risks of autonomous systems in critical workflows.

Although agents show strong potential for bug fixing and repository-level tasks, their use for microservice generation remains largely unexplored. Microservice  involve numerous fine-grained services communicating together ~\cite{dragoni2017microservices}, and development tools for supporting microservices are limited~\cite{vural2021ddd,zhong2024ddd}.
Microservices form a demanding testbed for AI agents because each service must satisfy its own logic while integrating through strict API contracts and interaction patterns. Even small errors can break cross-service communication, and existing tools still struggle with these complexities~\cite{dragoni2017microservices}. 
Generating microservices therefore requires both local correctness within each service and global correctness across service contracts, making this architectural style a realistic and challenging benchmark for assessing modern AI agent capabilities.

% https://anonymous.4open.science/r/microservices-4JF7K/
The \textbf{goal} of this study is to evaluate the functional correctness and code quality of AI agents-generated microservice. Our study employs 2 prompt types, 4 projects, and 3 AI agents, each generation was done in 2 scenarios and 3 different microservices per project producing 144 implementations evaluated for functionality, code quality, and efficiency across different generation scenarios\footnote{Code available at: \url{https://doi.org/10.5281/zenodo.17863951}}. We provide the following \textbf{main contributions}: 
\begin{itemize}
    \item The first empirical study on AI agents generating microservices under different contextual scenarios, evaluating functional correctness and code quality;
   \item A comparative analysis of agent efficiency, covering generation time, cost, and token consumption.
\end{itemize}

\noindent \textbf{Paper Structure}:  Section~\ref{sec:related} provides background information about AI Agents and related work. Section~\ref{sec:study} describes our research questions, the different Contextual Scenarios in our work  and the design of our study. Section~\ref{sec:results} presents results, which are discussed in Section~\ref{sec:discussion} along with threats to validity of our study. Finally, Section~\ref{sec:conclusion} presents our conclusion and future work.

% \section{Background ad Related Works}\label{sec:background}

% This section provides an overview of agentic AI systems for code generation and the contextual scenarios. 
% % that motivate our study.

% \subsection{Agentic AI for Code Generation}

% LLM-based agents extend the capabilities of standard Large Language Models (LLMs) with their autonomous task execution capabilities. While foundational LLMs operate as probabilistic engines that predict the next token based on a static input, agents such as OpenAI's Codex \cite{chen2021codex} wrap these models in a control loop that enables interaction with an external environment~\cite{dong2025survey}. Acode generation agent typically functions through an iterative perception-decision-execution cycle formalized as a trajectory of thoughts, actions, and observations \cite{bouzenia2025understandingsoftwareengineeringagents}. Within this cycle, the agent utilizes tools, such as file system interfaces and command-line executors to manipulate the environment and generate code. Specifically, the agent receives feedback from these actions (such as compilation errors or test logs etc.) allowing it to validate its assumptions and autonomously refine the implementation~\cite{yang2024sweagent, wu2025humanevalcomm}. This iterative capability is useful for generating code for microservices, which require precise adherence to file structures and API contracts that may not always be inferred from a single-pass generation.

\section{Background and Related Work}\label{sec:related}

This section provides an overview of agentic AI systems for code generation and describes related works
% that motivate our study.

\subsection{Agentic AI for Code Generation}

LLM-based agents extend the capabilities of standard Large Language Models (LLMs) with their autonomous task execution capabilities. While foundational LLMs operate as probabilistic engines that predict the next token based on a static input, agents such as OpenAI's Codex \cite{chen2021codex} wrap these models in a control loop that enables interaction with an external environment~\cite{dong2025survey}. Acode generation agent typically functions through an iterative perception-decision-execution cycle formalized as a trajectory of thoughts, actions, and observations \cite{bouzenia2025understandingsoftwareengineeringagents}. Within this cycle, the agent utilizes tools, such as file system interfaces and command-line executors to manipulate the environment and generate code. Specifically, the agent receives feedback from these actions (such as compilation errors or test logs etc.) allowing it to validate its assumptions and autonomously refine the implementation~\cite{yang2024sweagent, wu2025humanevalcomm}. This iterative capability is useful for generating code for microservices, which require precise adherence to file structures and API contracts that may not always be inferred from a single-pass generation.

\subsection{Related Works}
Over the years, promising work has been done in specifying software architecture using Architecture Description Languages (ADL) and supporting code generation through model transformations~\cite{brun2008code, malavolta2012industry}. For microservices specifically, Rademacher et al.~\cite{lemma-microservices} proposed an extensible approach to generate microservice code and deployment specifications from LEMMA models. Suljkanovi{\'c} et al.~\cite{silveria} proposed Silveria, a DSL that allows users to model microservice architectures and generate executable code using model transformations. However, as pointed out by Malavolta et al.~\cite{malavolta2012industry}, ADLs in general have a steep learning curve which often hinders practical adoption. Prior to the widespread adoption of LLMs, research on microservice development focused primarily on decomposition which algorithmically identifies optimal service boundaries within monolithic applications~\cite{microservice_survey}. These approaches employ heuristics based on static and dynamic analysis to cluster related components~\cite{decomposition_techniques}. These works however, are limited by the recommendation for service boundaries and not functional implementation.

Significant work has been done on using LLMs for code generation~\cite{bareiss2022codegen, gong2023intendedcodegen, chen2023improvingcodegen, wang2023review_llmcodegeneration, jiang2024surveylargelanguagemodels}. Recent work has shifted toward evaluating autonomous software engineering agents capable of complex, multi-step tasks~\cite{llm_agent_survey}. The SWE-Bench framework~\cite{swebench} has established test-pass rates as the primary evaluation metric for such agents, using "Fail-to-Pass" tests as the evaluation signal. Research has also evolved from simple Retrieval-Augmented Generation to iterative refinement frameworks~\cite{cocogen}, where agents generate code, receive compiler feedback, and refine their solutions. The formal analysis of agent trajectories, represented as sequences of (thought, action, result) triples~\cite{bouzenia2025understandingsoftwareengineeringagents}, enables researchers to analyze not just final outputs but the reasoning process.

% Remove if we are removing code quality
The quality of LLM-generated code remains debated. Some studies find LLM code more complex than human code~\cite{llm_code_smells}, while others observe low complexity~\cite{self_admitted_llm}. While other studies find~\cite{harmful_code}  that low complexity reflects missing defensive programming constructs like error handling and input validation. Additionally, benchmark contamination is a recognized methodological problem~\cite{contamination_survey}, as popular repositories are likely present in training corpora, potentially inflating performance on well-known benchmarks.

Recent work has demonstrated LLM-based generation of serverless functions as architectural components~\cite{arun}. We extend this to microservices, which are larger, stateful services with complex integration requirements. To our knowledge, no prior work has systematically evaluated autonomous agents for end-to-end microservice generation across varying contextual scenarios, measuring both functional correctness through automated testing and code quality through established metrics.

\begin{figure*}[tb]
    \centering
    \includegraphics[width=1.0\linewidth]{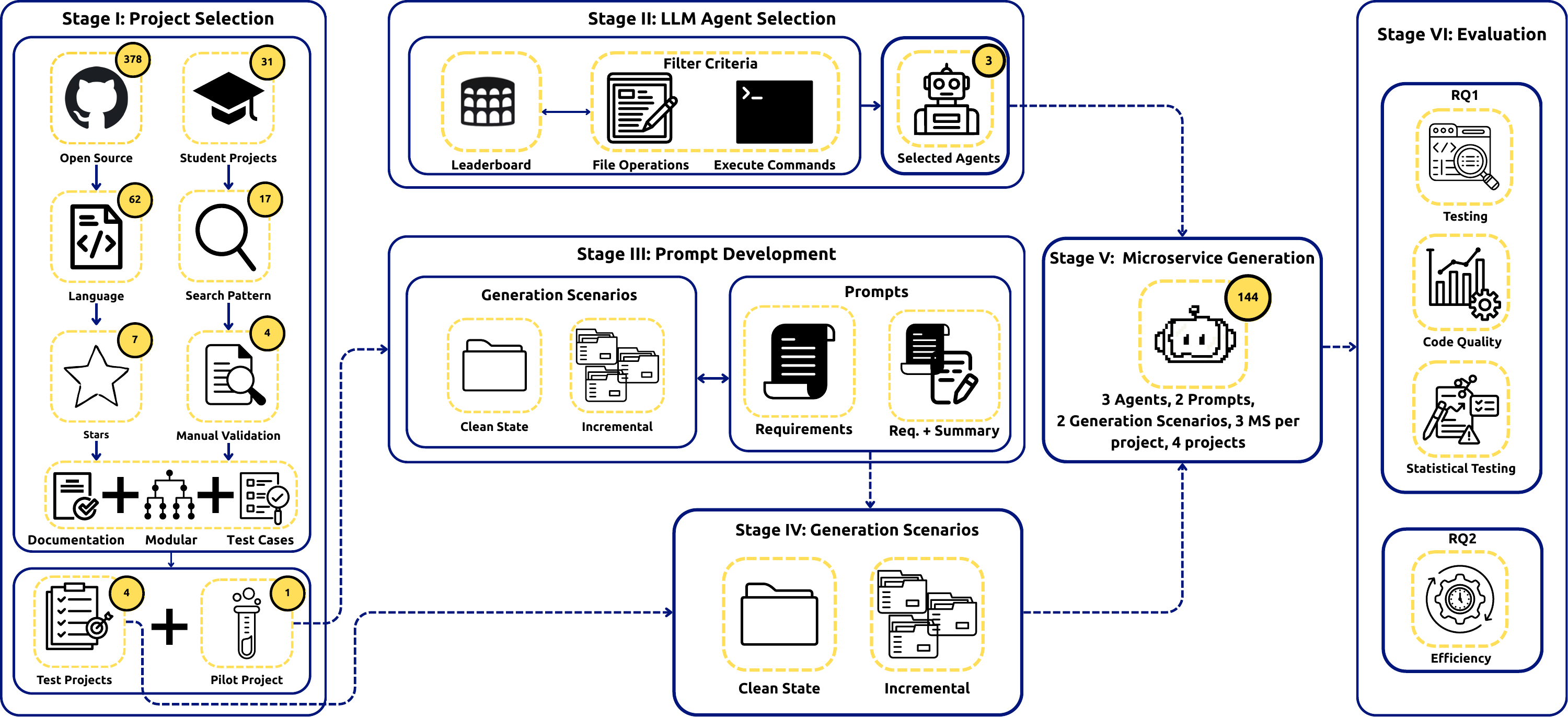}
    \caption{Overview of the Study Design}
    \label{fig:study-design}
\end{figure*}

\section{Study Design}\label{sec:study}
\subsection{Goal}
This study aims to evaluate the generation capabilities of AI agents in microservice development across varying levels of contextual information. Using the Goal-Question-Metric approach \cite{caldiera1994goal}, we formalize our goal to:

% The GQM template follows this structure:
% Analyze [object of study] for the purpose of [purpose] with respect to [quality focus] from the point of view of [perspective] in the context of [environment].
% Should the GQM change? considering we don't mention efficiency here. edit: added efficiency 
\textbf{analyze} the effectiveness of AI agents \textbf{for the purpose of} generating microservices \textbf{with respect to} functional correctness, code quality and efficiency \textbf{from the viewpoint of} software developers and architects \textbf{in the context of} existing microservice-based systems.
\subsection{Research Questions}

% \noindent \textbf{RQ$_{1}$: To what extent can AI agents generate functionally correct and low complexity microservice code across different contextual scenarios?}

\noindent \textbf{RQ$_{1}$: To what extent can AI agents generate microservices across different contextual scenarios, in terms of functional correctness and  code quality?}

\begin{itemize}
    \item \textbf{RQ$_{1.1}$}: To what extent can AI agents develop microservices in \textbf{incremental} generation scenarios?
    \item \textbf{RQ$_{1.2}$}: To what extent can AI agents develop microservices in \textbf{clean state} generation scenarios?
\end{itemize}

% \\\\\\\\
% Option 1
% In the incremental generation scenario (defined in Section \ref{sec:scenarios}), we assess whether AI-generated microservices are functional when the agents receive comprehensive contextual information, including the existing system implementation, documentation, and preserved API integration points (RQ$_{1.1}$).

% Matyteo
In the incremental generation scenario,
%During \textbf{Incremental} generation, a team adds a microservice to an already existing system with already running services, established patterns, documented API contracts, and provided templates for configuration. Here, the main challenge is to integrate correctly into the existing system while respecting API contracts, similar to adding functionality to established ecosystems.  
We investigate whether AI agents can generate microservices based on requirements specifications, and comprehensive contextual information, including the existing system implementation, documentation, and preserved API integration points (\textbf{RQ$1.1$}). This mirrors the real-world scenario where a team adds a microservice to an already existing system with already running services, established patterns, documented API contracts, and provided templates for configuration. 
 % Here, the main challenge is to integrate correctly into the existing system while respecting API contracts, similar to adding functionality to established ecosystems.
% . Thereby, agents must infer system architecture and implementation details from minimal contextual information (\textbf{RQ$1.2$}). 

In clean state generation, we investigate whether AI agents can generate microservices based only on the requirements specifications, without any information of the system implementation (\textbf{RQ$1.2$}). This is similar to the real-world case, which involves teams working from requirements alone on \textbf{greenfield projects} or \textbf{early-stage development}. Available information \textbf{only includes} requirements specifications. Inferring architectural patterns, component interactions, and API designs from minimal context is challenging. This involves significantly more exploration and decision-making compared to incremental generation, where agents can study existing patterns and adapt to them.

We measure \textbf{functional correctness} through automated test execution and assess \textbf{code quality} through standard software metrics including size (SLOC), complexity (Cyclomatic Complexity), and understandability (Cognitive Complexity). 

% Option 2
% For incremental generation (see Section \ref{sec:scenarios}), we examine whether AI agents can produce functional microservices when given full contextual information, such as the current system implementation, documentation, and the existing API integration points (RQ$_{1.1}$)/
% \\\\\\\\\

% In incremental generation (defined in \ref{sec:scenarios}), we evaluate whether microservices generated by AI agents are functional when provided with comprehensive contextual information including existing system implementations, documentation, and preserved API integration points \textbf{(RQ$_{1.1}$)}. 

\noindent  \textbf{RQ$_{2}$: How do AI agents compare in efficiency when generating microservices?}

In this RQ, we compare the resource efficiency of different AI agents by measuring the computational cost (token consumption and associated monetary costs) and time required for generation. This analysis helps determine the practical feasibility and relative performance of using different AI agents for microservice development in real-world scenarios.
% ---------------------------------------------------------

\subsection{Experiment Workflow}

\subsubsection{Project Selection}

% \begin{figure}[tb]
% \centering
% \includegraphics[width=1.0\linewidth, height=1.0\textheight, keepaspectratio]{images/funnel.pdf}
% \caption{Project Selection Process}
% \label{fig:main-overview}
% \end{figure}

% Okay, removing this for now
% \todo[inline]{Matteo: Fig. \ref{fig:main-overview}is hard to read}

We systematically collected microservice systems from two primary sources to ensure diversity.  Our data collection process follows a multi-stage filtering approach to identify suitable repositories for LLM agent evaluation.

\textbf{Open Source Repository Selection:} We began with an existing microservice dataset \cite{2024davidedataset} consisting of 378 sample points and applied a filter based on the patterns seen in Java-based Spring Boot applications, this choice of language was based on familiarity and proper build mechanism. We chose repositories which have over 40\% Java content to include repositories that may contain web frontend code (HTML, CSS, JavaScript) as part of a mono-repository structure while having sufficient Java-based microservice backend code for our evaluation, using GitHub API. This gave us 62  repositories. We then filtered repositories based on popularity metrics, requiring systems to have more than 600 GitHub stars to ensure real-world relevance and community validation. This filtering process yielded 7 repositories, from which we further refined our selection to 2 repositories that met our technical requirements and were also widely used in microservice literature \cite{2024fischerMicroservicesTesting, 2020RegressionMicroservices}. We required systems to provide test cases for functional correctness validation, documentation, and detailed requirements specifications.

\noindent \textbf{Private Project Integration:} To complement open source systems which might have been subject to LLM contamination \cite{contamination_survey}, we incorporated private projects from software engineering coursework. In this coursework, students were asked to implement a software system with an architecture of their choice and to provide proper documentation specifying requirements, architectural viewpoints and diagrams. They were also asked to implement an alternative architecture and provide comparison of performance, state pros-cons. Out of the 31 projects we filtered the term "microservice" (or its alternatives "micro-service"/"microservices" etc.) in the document, shortlisting 17 candidate projects. On investigation, 4 projects had used the microservice architecture as their main implementation and not as an alternative, these are the ones we selected for further examination. We manually verify that these projects contain documentation and have a modular code structure before finally selecting 3 private projects (2 for evaluation and 1 for pilot study).

\textbf{Final Dataset Composition:} Our complete filtering pipeline resulted in 5 microservice systems with varying levels of complexity. Table~\ref{tab:microservice-overview} presents the  characteristics of the projects.

\begin{table}[!htbp]
\centering
\caption{Selected Microservice Projects (O=Open Source, P=Private)}
\label{tab:microservice-overview}
\renewcommand{\arraystretch}{1.3}
\begin{tabular}{p{1.5cm}lcp{3.8cm}}
\hline
\textbf{Project} & \textbf{Lang.} & \textbf{Services} & \textbf{Description} \\
\hline
PiggyMetrics (O) & Java & 3 & Personal finance management with account statistics and notifications \\
Train-Ticket (O) & Java & 41 & Train ticket booking with search, orders, payment, and user management \\
TeamSync (P) & Java & 4 & Team collaboration platform for course registration and task management \\
Project Management (P) & Python & 4 & Collaborative task management with hierarchical task breakdown \\
MicroBank (P, Pilot) & Python & 4 & Banking system with account management and transaction processing \\
\hline
% \vspace{-2}
\end{tabular}%
\end{table}

\subsubsection{LLM Agent Selection}

To evaluate microservice generation capabilities We selected coding agents using established benchmarks while prioritizing both
state-of-the-art performance and open-source representation. We select
agents from the top performers of the LMArena Coding  Leaderboard\footnote{\url{https://lmarena.ai/}} \cite{chiang2024chatbotarenaopenplatform}.
Our selection prioritized LLMs with available coding agents to  perform iterative file operations, execute bash commands, and interact with codebases through tool-calling as microservice generation requires system-level interactions beyond simple code completion. We also use the default coding agents provided by vendors of the LLMs selected, with their default settings and prompts.

The agents selected for microservice generation are described in  Table~\ref{tab:agent-info} below. "Access Method" indicates whether the agent requires a paid subscription (Subscription), API token access (API), or can be  hosted locally (Local). For agents offering multiple access methods, we  highlight the method used in our study in bold. "License Type" refers to whether the model weights are publicly available (Open) or not (Proprietary). All experiments were conducted on a local machine using these coding agents in non-interactive mode with permission bypass flags for uninterrupted execution. We perform the efficiency analysis through the logs of different experimental runs.

\begin{table}[!htbp]
\centering
\caption{Characteristics of Selected LLM Coding Agents}
\label{tab:agent-info}
\renewcommand{\arraystretch}{1.3}

\begin{tabular}{p{1cm}lllp{1cm}}
\hline
\textbf{Agent} & \textbf{Model} & \textbf{Context} & \textbf{Access} & \textbf{License} \\
\textbf{Name} & & \textbf{Window} & \textbf{Method} & \textbf{Type} \\
\hline
Claude Code & claude-sonnet-4 & 1M & \textbf{Subscription}/ & Proprietary \\
 & -20250514 &  & API & \\
Codex & GPT-5 & 400K & \textbf{Subscription}/ & Proprietary \\
 & (high effort) &  & API & \\
Code Qwen & qwen3-coder & 1M & \textbf{API}/Local & Open \\
 & -480b-a35b &  &  & \\
\hline
\end{tabular}%
\end{table}

\subsubsection{Pilot Study and Prompt Development}

Before conducting the full evaluation, we refined our prompt design through a pilot study using our pilot project with Claude Code. Through this process, we developed a two-stage approach where we generate implementation summaries of existing services using the prompt shown in \ref{box:summary_prompt}. Further, we modified our prompts to include role personification, framing agents as senior engineers. We observed that the generation across multiple runs had similar outputs with this pilot study.

\begin{tcolorbox}[colback=orange!5!white, colframe=orange!95!white,  breakable = true,colbacktitle=orange!95!white, title=Summary Generation Prompt Template, label={lst:summary_gen}]
\small
    \textbf{Role:} Expert software architect with 15+ years of experience in microservices architecture and enterprise system design, specializing in analyzing complex codebases and distilling them into clear descriptions.

    \tcblower
\small
    \textbf{Task:} Analyze \textit{\{MICROSERVICE\_NAME\}} and generate a comprehensive description capturing its complete implementation for regeneration purposes.
    
    \textbf{Instructions:} Read requirements document \textit{\{REQUIREMENTS\_PATH\}}, explore codebase to identify all related files, cross-reference implementation with requirements, focus only on this specific microservice (ignore others except interaction points).
    
    \textbf{Output Format:} Generate high-level description including:
    \begin{itemize}
        \item Functional Scope (core responsibilities)
        \item Key Features Implemented
        \item API Responsibilities
        \item Data Responsibilities (managed data and business rules)
        \item Integrations (connections with other services from this service's perspective)
        \item Technical Approach (high-level decisions and patterns)
        \item Implementation Notes (business logic and special considerations)
    \end{itemize}
\normalsize
\end{tcolorbox}
\begin{minipage}{\linewidth}
\label{box:summary_prompt}
\end{minipage}

\textbf{Prompt Strategies:} For each experiment type, we employ two prompt strategies:

\begin{itemize}
    \item \textit{P1 (Name + Requirements Path)}: Provides only the microservice name and path to requirements document, representing minimal contextual information. Agents must explore the codebase in Incremental or work from requirements alone in Clean State to understand what needs to be implemented.
\end{itemize}
    \begin{tcolorbox}[colback=orange!5!white, colframe=orange!95!white, breakable = true,colbacktitle=orange!95!white, title=Minimal Context Prompt (P1), width=\linewidth, label={lst:p1_prompt}]
\small
    \textbf{Role:} Senior software engineer with 12+ years experience in distributed systems and microservices architecture.

    \tcblower
\small
    \textbf{Task:} Generate complete microservice implementation.
    
    \textbf{Target:} \textit{\{MICROSERVICE\_NAME\}}
    
    \textbf{Requirements:} \textit{\{REQUIREMENTS\_PATH\}}
    
    \textbf{Instructions:} Explore codebase, understand patterns and tech stack, generate complete implementation with all components, verify compilation.
    
\normalsize
\end{tcolorbox}
\begin{minipage}{\linewidth}
\label{box:p1_prompt}
\end{minipage}

% \begin{table}[!htbp]
% \centering
% \caption{Complexity Metrics Comparison of Microservices Projects}
% \label{tab:complexity_comparison}
% \resizebox{\columnwidth}{!}{%
% \begin{tabular}{llrrrr}
% \toprule
% \textbf{Project} & \textbf{Lang.} & \textbf{MS} & \textbf{LoC} & \textbf{CycC} & \textbf{CogC} \\
% \midrule
% Train Ticket & Java & 41 & 18,203 & 1,841 & 1,258 \\
% PiggyMetrics & Java & 3 & 1,927 & 235 & 33 \\
% TeamSync & Java & 4 & 1,942 & 295 & 144 \\
% CLI Task Mgr & Python & 4 & 7,716 & 1,294 & 1,150 \\
% \bottomrule
% \end{tabular}%
% }
\begin{itemize}
    \item \textit{P2 (Name + Requirements Path + Implementation Summary)}: Includes microservice name, path to requirements document, and a generated implementation summary of the target microservice from \ref{box:summary_prompt}, capturing the microservice's functional scope, API responsibilities, data management patterns, integration points with other services, technical approach, and implementation notes. Since the requirements contains details about the entire system, the summary highlights the information of the target microservice to be generated to assist the agent.
\end{itemize}

\begin{tcolorbox}[colback=orange!5!white, colframe=orange!95!white, breakable = true,colbacktitle=orange!95!white, title=Prompt with Implementation Summary (P2), width=\linewidth, label={lst:p2_prompt}]
\small
    \textbf{Role:} Senior software engineer with 12+ years experience in distributed systems and microservices architecture.

    \tcblower
\small
    \textbf{Task:} Generate complete microservice implementation.
    
    \textbf{Target:} \textit{\{MICROSERVICE\_NAME\}}
    
    \textbf{Requirements:} \textit{\{REQUIREMENTS\_PATH\}}
    
    \textbf{Summary:} \textit{\{GENERATED\_SUMMARY from \ref{lst:summary_gen}: Functional scope, API/data responsibilities, integrations, technical approach, implementation notes\}}
    
    \textbf{Instructions:} Explore codebase, follow summary guidance, generate from scratch, verify compilation.
    
\normalsize
\end{tcolorbox}
\begin{minipage}{\linewidth}
\label{box:p2_prompt}
\end{minipage}

\subsubsection{Generation Scenarios}

\textbf{Contextual Scenarios:} 

%In real-world development, the available context during the creation of a microservice may vary. During \textbf{Incremental} generation, a team adds a microservice to an already existing system with already running services, established patterns, documented API contracts, and provided templates for configuration. Here, the main challenge is to integrate correctly into the existing system while respecting API contracts, similar to adding functionality to established ecosystems.  \textbf{Clean State} generation involves teams working from requirements alone on greenfield projects or early-stage development. Available information only includes requirements specifications. Inferring architectural patterns, component interactions, and API designs from minimal context is challenging. This involves significantly more exploration and decision-making compared to incremental generation where agents can study existing patterns and adapt to them.
We evaluate agents across two generation scenarios representing different levels of system context availability:

\begin{itemize}
    \item \textit{Incremental Generation (RQ$_{1.1}$)}: The target microservice's code is removed from the codebase, but all integration points and dependencies remain intact, allowing agents to explore the existing architecture before generation. This is done by removing the directory of the microservice since we used mono-repository modular projects. Agents have access to the test cases.

    \item \textit{Clean State Generation (RQ$_{1.2}$)}: The target microservice implementation traces are removed from the system, including the directory, configuration files, and API calls made to the microservice from other microservices. This is done by locating API calls to external services using search patterns or by referring to documentation for architectural details. Agents must infer system architecture and implementation details from requirements specifications alone without access to existing service implementations. 
\end{itemize}
The original code acts as a baseline to compare metrics for code quality assessment.

\subsubsection{Microservice Generation}

Each experiment type and prompt strategy combination represents a distinct experimental condition, resulting in four conditions total: Inc-P1, Inc-P2, CS-P1, and CS-P2. For each project, we selected three microservices based on their business logic importance and removed them one at a time for generation. For each condition, we evaluate all three selected agents across our four microservice projects, with each project contributing three generation tasks, generating a total of 144 microservice implementations (4 conditions × 3 agents × 4 projects × 3 microservices per project).

\subsubsection{Evaluation Metrics}

For code quality analysis, we used SonarQube throughout the projects, test execution was automated using project-specific testing frameworks: Spring Boot Test for Java projects and unit test for Python projects.

\textbf{Functional Correctness (RQ$_{1.1}$ and RQ$_{1.2}$):} We evaluated functional correctness through automated testing, with different strategies for each experiment type:

\begin{itemize}
    \item \textit{Incremental Generation (RQ$_{1.1}$)}: We executed the original unit test suites associated with each target microservice. Agents had access to these test cases during generation. We calculated test pass rates by dividing passing tests by total tests for the individual microservices.

    \item \textit{Clean State Generation (RQ$_{1.2}$)}: We assessed integration success by executing tests from other microservices that consume the generated service's APIs. This approach was necessary because agents generating from minimal context often produced functionally correct code with different package structures, class names, and architectural patterns than expected by baseline unit tests. Figure~\ref{fig:cleanstate-mismatch} illustrates a representative example from the Train-Ticket payment service, where Codex generated correct payment functionality but the syntactical mismatch causes compilation failures in unit tests despite functional correctness. We measured success by determining whether dependent microservices tests passed without failures.
\end{itemize}

\begin{figure}[tb]
\centering
{\small \textbf{Expected Structure (from baseline unit tests):}}
\begin{lstlisting}
package com.trainticket.service;

import com.trainticket.entity.Money;
import com.trainticket.entity.Payment;
import com.trainticket.repository.AddMoneyRepository;
import com.trainticket.repository.PaymentRepository;

public class PaymentServiceImplTest {
    @Mock
    private PaymentRepository paymentRepository;
    @Mock
    private AddMoneyRepository addMoneyRepository;
}
\end{lstlisting}

\vspace{0.2cm}
{\small \textbf{Codex Generated Implementation:}}
\begin{lstlisting}
// File: payment/entity/MoneyTransaction.java
public class MoneyTransaction { ... }

// File: payment/repository/MoneyTransactionRepository.java
public interface MoneyTransactionRepository { ... }
// Note: Money and AddMoneyRepository classes not generated
\end{lstlisting}

\caption{Example of structural mismatch in Clean State generation (Train-Ticket payment service): baseline tests expect \texttt{com.trainticket.*} package with \texttt{Money} and \texttt{AddMoneyRepository} classes, but Codex generated \texttt{payment.*} package with \texttt{MoneyTransaction} and \texttt{MoneyTransactionRepository}, causing compilation failures despite functional correctness}
\label{fig:cleanstate-mismatch}
\end{figure}

\textbf{Code Quality Metrics (RQ$_{1.1}$ and RQ$_{1.2}$):} We quantify code quality using metrics that measure size and complexity, comparing generated code against human-written baselines:

\begin{itemize}
    \item \textit{Source Lines of Code (SLOC)}: Quantifies program size by counting lines containing source code.
    \item \textit{Cyclomatic Complexity (CC)} \cite{1976cyclomatic}: Measures control flow complexity and testability.
    \item \textit{Cognitive Complexity (CogC)} \cite{2018cognitive}: Measures code understandability.
\end{itemize}

For each metric, we calculated the change from baseline across experimental configurations, agents, and projects. We use SonarQube to uniformly assess these metrics, as alternative approaches like CodeBLEU \cite{ren2020codebleu} and CodeBERTScore \cite{zhou2023codebertscore} are designed primarily for function-level snippets rather than full projects.

\textbf{Efficiency Metrics (RQ$_{2}$):} We measured resource consumption and efficiency:

\begin{itemize}
    \item \textit{Token Consumption}: Total input and output tokens.
    \item \textit{Cost}: Associated monetary costs based on each agent's pricing model, using API pricing.
    \item \textit{Generation Time}: Total time from generation start to completion in minutes.
\end{itemize}

We use the logs provided by the generation runs to get cost based on the tokens and perform all the experiments under similar network setting.

\textbf{Statistical Analysis:} We first assessed whether the data were normally distributed using the Anderson-Darling (AD) test \cite{arcuri2011practical}, which evaluates whether a set of observations follows a specified probability distribution by measuring the differences between the empirical cumulative distribution function of the data and the theoretical distribution. Results of the statistical test (A\textsuperscript{2} = 20.068, p $\lt$ 0.001) allowed us to reject the null hypothesis of normality, confirming that the gathered metrics do not follow a normal distribution.

After establishing non-normality, we employed the Wilcoxon signed-rank \cite{yi2025experimental} test to compare test pass rates across prompt strategies (P1 vs P2). The Wilcoxon signed-rank test is a non-parametric statistical hypothesis test that compares two related samples or repeated measurements to assess whether their population mean ranks differ. We further assessed the magnitude and direction of statistically significant differences using the Dunn-All Pair test \cite{liu2021exploratory}, which performs pairwise comparisons while controlling for multiple testing. We set the significance level ($\alpha$) to 0.01, lowering it from the conventional 0.05 to account for the large number of statistical tests performed.

\section{Results}\label{sec:results}
% \todo[inline]{Matteo: we should increase font size ..., B: plz check now}
% \begin{itemize}
%     \item \textbf{RQ$_{1.1}$: To what extent can AI agents develop microservices in incremental generation scenarios?}

%     \item \textbf{RQ$_{1.2}$: To what extent can AI agents develop microservices in clean state generation scenarios?}
% \end{itemize}

\subsection{RQ$_{1.1}$: To what extent can AI agents develop microservices in incremental generation scenarios?}
\textbf{Functional Correctness:} We analyze the test pass rates against the Baseline (B) pass rates of the original project,  for incremental generation, where agents had access to the existing system architecture and integration points (Table~\ref{tab:rq1_effectiveness}). 
Codex achieved the highest average performance with P1 prompts (75.9\%), followed by Claude Code (73.7\%) and Code Qwen (50.5\%). Providing additional context through P2 prompts and generating the target microservice did not consistently improve performance, with average test pass rates decreasing to 50.3\% for Codex, 63.2\% for Claude Code, and 47.2\% for Code Qwen. In this setting, Code Qwen had comparable results to the other agents. Further,
\begin{table}[tb]
    \centering
    \scriptsize
    \renewcommand{\arraystretch}{1.2}
    \setlength{\tabcolsep}{3pt}
    \caption{Test Pass Rates for Incremental and Clean State Development for Different Types of Prompts P1 and P2}
    \label{tab:rq1_effectiveness}
    \begin{tabular}{llcccccc}
        \hline
        \multirow{3}{*}{\textbf{Agent}} & \multirow{3}{*}{\textbf{Project}} & \multicolumn{6}{c}{\textbf{Test Pass Rate (\%)}} \\
        \cline{3-8}
        & & \multicolumn{3}{c}{\textbf{Incremental}} & \multicolumn{3}{c}{\textbf{Clean State}} \\
        \cline{3-8}
        & & \textbf{B} & \textbf{P1} & \textbf{P2} & \textbf{B} & \textbf{P1} & \textbf{P2} \\
        \hline
        \multirow{5}{*}{Claude Code}
        & Piggymetrics & 100.0 & 91.1 & 81.7 & 100.0 & 98.0 & 98.0 \\
        \cline{2-8}
        & Train-Ticket & 100.0 & 92.6 & 60.6 & 81.5 & 97.0 & 97.0 \\
        \cline{2-8}
        & TeamSync & 100.0 & 66.7 & 66.7 & 100.0 & 100.0 & 100.0 \\
        \cline{2-8}
        & Project-Management & 92.3 & 44.3 & 43.9 & 92.3 & 92.6 & 96.3 \\
        \cline{2-8}
        & \textbf{Avg. Agent Performance} & \textbf{-} & \textbf{73.7} & \textbf{63.2} & \textbf{-} & \textbf{96.9} & \textbf{97.8} \\
        \hline
        \multirow{5}{*}{Codex}
        & Piggymetrics & 100.0 & 93.8 & 93.3 & 100.0 & 99.0 & 65.7 \\
        \cline{2-8}
        & Train-Ticket & 100.0 & 100.0 & 59.8 & 81.5 & 97.0 & 97.0 \\
        \cline{2-8}
        & TeamSync & 100.0 & 66.7 & 0.0 & 100.0 & 100.0 & 100.0 \\
        \cline{2-8}
        & Project-Management & 92.3 & 43.0 & 48.0 & 92.3 & 96.3 & 63.0 \\
        \cline{2-8}
        & \textbf{Avg. Agent Performance} & \textbf{-} & \textbf{75.9} & \textbf{50.3} & \textbf{-} & \textbf{98.1} & \textbf{81.4} \\
        \hline
        \multirow{5}{*}{Code Qwen}
        & Piggymetrics & 100.0 & 66.7 & 84.1 & 100.0 & 98.0 & 98.0 \\
        \cline{2-8}
        & Train-Ticket & 100.0 & 60.6 & 25.3 & 81.5 & 33.3 & 97.0 \\
        \cline{2-8}
        & TeamSync & 100.0 & 33.3 & 33.3 & 100.0 & 100.0 & 100.0 \\
        \cline{2-8}
        & Project-Management & 92.3 & 41.5 & 46.0 & 92.3 & 96.3 & 63.0 \\
        \cline{2-8}
        & \textbf{Avg. Agent Performance} & \textbf{-} & \textbf{50.5} & \textbf{47.2} & \textbf{-} & \textbf{81.9} & \textbf{89.5} \\
        \hline
    \end{tabular}
\end{table}
Table~\ref{tab:rq1_effectiveness} reveals that commercial agents, namely Codex and Claude Code, outperformed the open-source alternative Code Qwen, regardless of prompt type. Notably, Codex achieved 100\% test pass rate for Train-Ticket (the largest system among the candidates) with P1 prompts, while Code Qwen struggled particularly with TeamSync and Train-Ticket, achieving upto 33.3\% pass rates in some configurations.

\begin{figure*}[tb]
    \centering
    \includegraphics[width=0.80\textwidth]{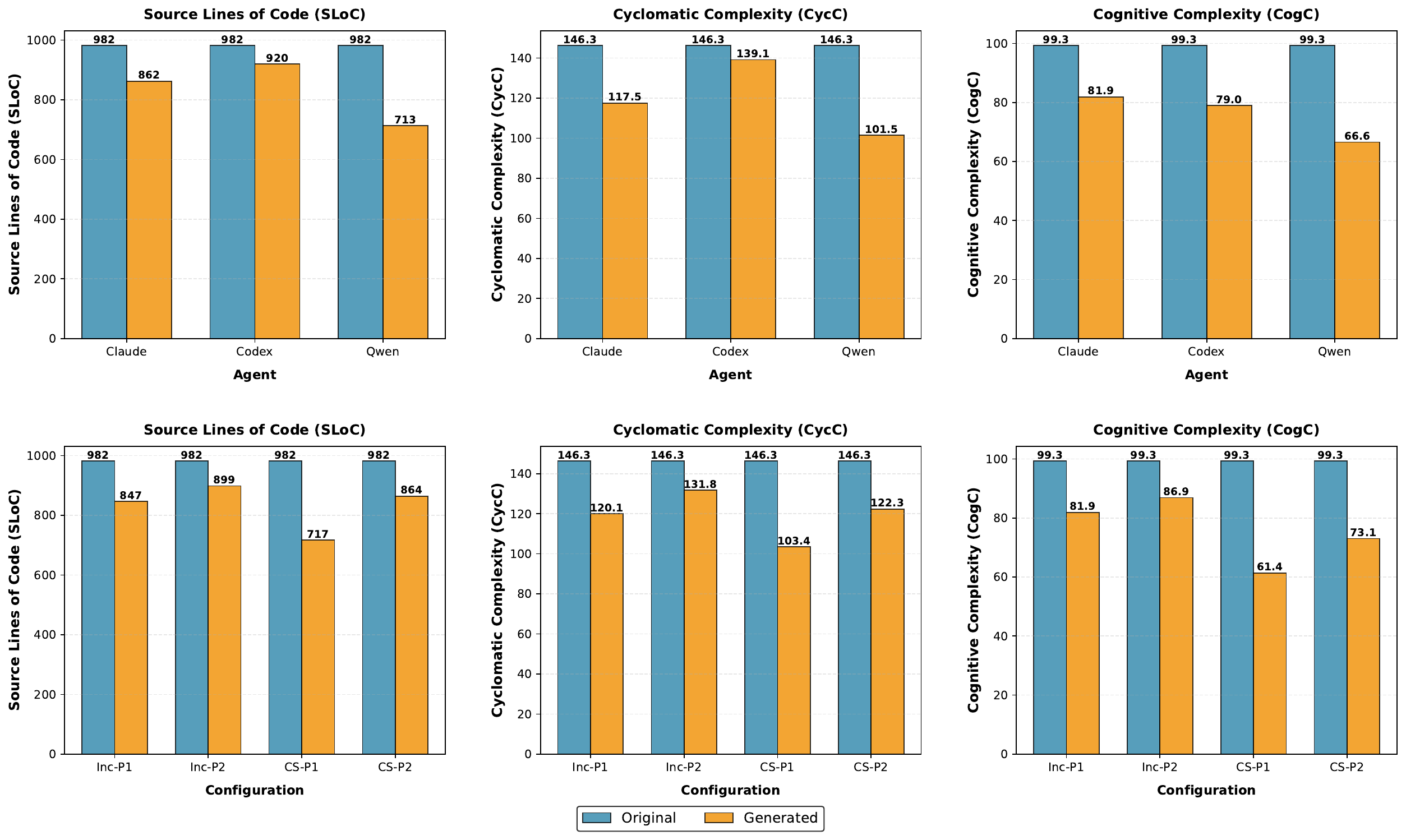}
    \caption{Code Quality Metrics comparison: Lines of Code (LoC), Cyclomatic Complexity (CycC), and Cognitive Complexity (CogC). Top row shows comparison by agent, bottom row shows comparison by configuration.}
    \label{fig:code_quality_clean}
\end{figure*}

\textbf{Code Quality.} To assess the quality of generated code beyond functional correctness, we analyzed static code metrics including Lines of Code (LoC), Cyclomatic Complexity (CycC), and Cognitive Complexity (CogC) using SonarQube. Figure~\ref{fig:code_quality_clean} presents a comparison of original versus generated code across agents and configurations.

The results show that generated code generally exhibits lower complexity metrics compared to baseline implementations. As shown in the top row of Figure~\ref{fig:code_quality_clean}, Claude Code and Codex consistently produced code with 20-30\% lower cyclomatic and cognitive complexity across all configurations, while maintaining comparable LoC. Code Qwen showed higher variance, particularly in the incremental P2 configuration where complexity metrics approached baseline levels.

The bottom row of Figure~\ref{fig:code_quality_clean} presents configuration-level comparisons, revealing that code quality improvements were consistent across different prompt and generation strategies, with all agents demonstrating reductions in complexity across most configurations. 

\subsection{RQ$_{1.2}$: To what extent can AI agents develop microservices in clean state generation scenarios?}

\textbf{Functional Correctness.} The clean state generation scenario displays different performance patterns. Agents achieved substantially higher average test pass rates: Claude Code (97.8\% with P2), Codex (98.1\% with P1), and Code Qwen (89.5\% with P2) (Table~\ref{tab:rq1_effectiveness}). Claude Code had the most consistent performance, even achieving 100\% integration success for TeamSync regardless of prompt strategy, and maintained 97-98\% success rates for Piggymetrics and Train-Ticket. Codex achieved the highest overall performance with P1 prompts (98.1\% average), but showed inconsistent results with P2 prompts, particularly dropping to 65.7\% for Piggymetrics and 63.0\% for Project-Management. Different from other agents, Code Qwen failed to write any code in 3 instances with P1, where it just terminated after reading files and realizing its in a loop (for eg- in generating the preserve microservice for Train-Ticket). However, when provided with implementation summaries through P2 prompts, Code Qwen recovered its performance, achieving 97.0\% integration success, a 63.7 percentage point improvement. All agents achieved perfect 100\% integration success for TeamSync. 

\textbf{Code Quality.} The code quality analysis demonstrates that clean state development generally produces simpler code across all agents, with reductions in cyclomatic and cognitive complexity ranging from 15-40\% below baseline (Figure~\ref{fig:code_quality_clean}---denoted by CS-P1, CS-P2). CS-P1 configurations consistently show the largest reductions in code complexity. Claude Code maintained the most consistent quality across both development approaches, while Code Qwen showed variability in the clean state scenarios.
 
% \vspace{-50pt}
\textit{Statistical Significance:} To validate observed performance differences, we conducted statistical analysis using non-parametric tests with significance level $\alpha$ = 0.01. The Wilcoxon signed-rank test revealed that the performance difference between incremental and clean state generation scenarios is highly significant (Z = 5.180, p $\lt$ 0.001). The Dunn-All Pair post-hoc test confirmed this finding, showing clean state achieves a mean difference of 34 percentage points higher test pass rates compared to incremental (Z = 5.180, p $\lt$ 0.001). This confirms that the choice of generation scenario (RQ$_{1.1}$ vs RQ$_{1.2}$) has a statistically significant impact on functional correctness. In contrast, prompt type differences (P1 vs P2) did not show statistical significance (Z = -1.475, p = 0.140 $\gt$ 0.01), indicating that while numerical variations exist across prompt strategies, these differences could be attributed to random variation or agent-specific characteristics rather than systematic prompt effects.

\begin{tcolorbox}[colback=orange!5!white, colframe=orange!95!white, colbacktitle=orange!95!white]
\small
    \textbf{Main Findings for RQ$_{1}$}: AI agents can generate functional microservices, but performance varies significantly by context. Incremental generation benefits from minimal prompts that encourage exploration, while clean state generation achieves high pass rates due to integration test flexibility. In clean state scenarios, explicit guidance can help overcome the limitations of the underlying model used. Human oversight remains necessary to ensure API contract compliance.
\end{tcolorbox}

\subsection{RQ$_{2}$: How do AI agents compare in efficiency when generating microservices?}

To compare the efficiency of different AI agents for microservice development, we analyze the generation time, computational cost, and token consumption. Table~\ref{tab:rq2_efficiency_summary} presents aggregate statistics across all 48 runs per agent (4 projects × 3 services × 4 configurations).

\begin{table}[!htbp]
    \centering
    \small
    \renewcommand{\arraystretch}{1.3}
    \setlength{\tabcolsep}{4pt}
    \caption{Average Efficiency Metrics Summary Across All Configurations (48 runs per agent)}
    \label{tab:rq2_efficiency_summary}
    \begin{tabular}{lcccccc}
        \hline
        \textbf{Agent} & \textbf{Avg} & \textbf{Total} & \textbf{ Input} & \textbf{Output} & \textbf{Avg} & \textbf{Total} \\
        & \textbf{Time} & \textbf{Time} & \textbf{Tokens} & \textbf{Tokens} & \textbf{Cost} & \textbf{Cost} \\
        & \textbf{(min)} & \textbf{(hrs)} & \textbf{(K)} & \textbf{(K)} & \textbf{(\$)} & \textbf{(\$)} \\
        \hline
        Claude Code & 7.8 & 6.20 & 4,416 & 2.1 & 13.28 & 637.39 \\
        Codex & 16.6 & 13.25 & 4,436 & 37.5 & 5.92 & 284.19 \\
        Code Qwen & 7.6 & 6.10 & 2,916 & 16.2 & 2.98 & 143.08 \\
        \hline
    \end{tabular}
\end{table}

\subsubsection{Time Efficiency}

\begin{figure}[tb]
    \centering
    \includegraphics[width=0.8\linewidth]{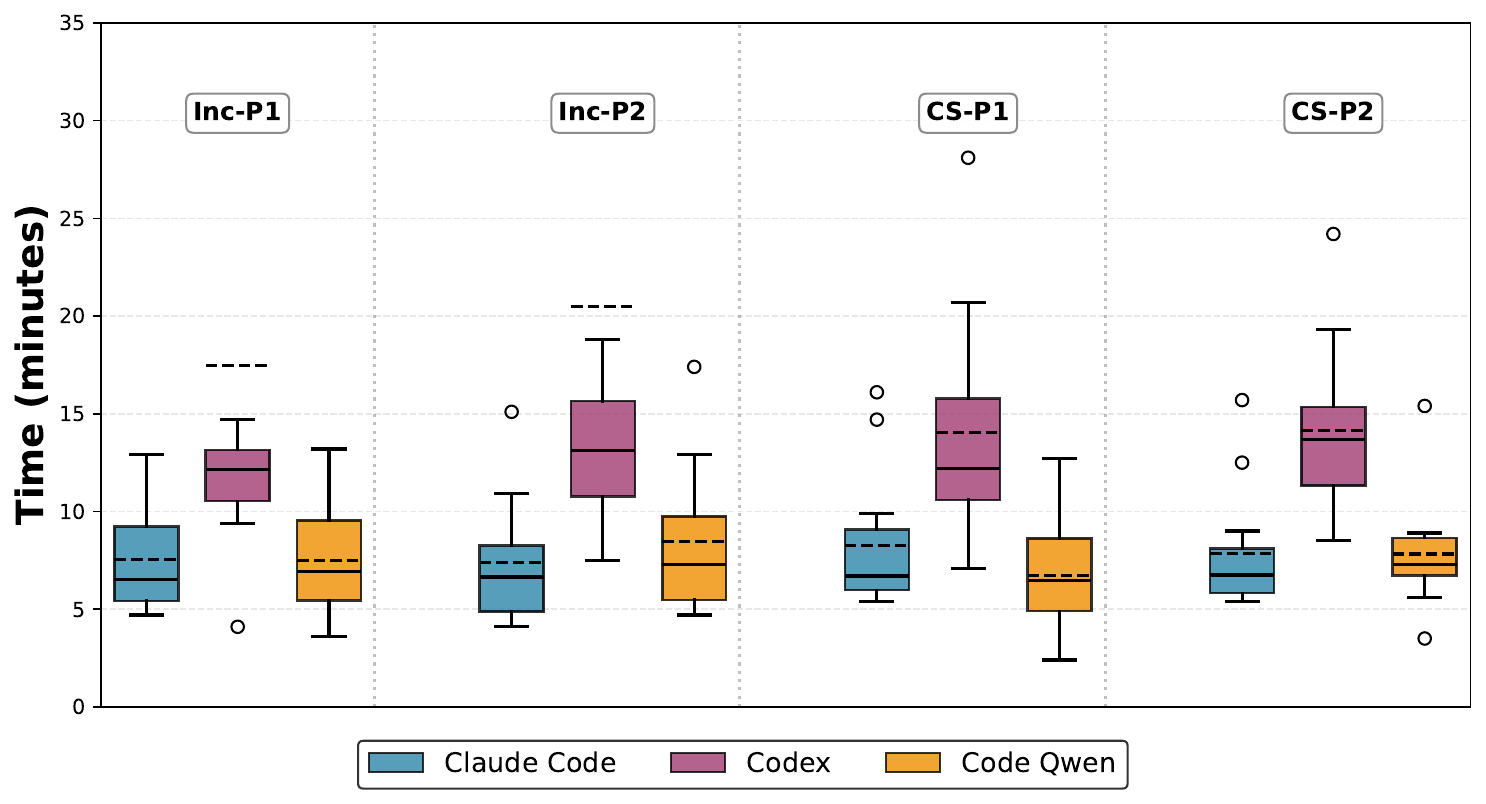}
    \caption{Time distribution across agents and configurations}
    \label{fig:efficiency_time}
    \vspace{-7pt}
\end{figure}

Figure~\ref{fig:efficiency_time} shows the distribution of generation times across all configurations. Claude Code and Code Qwen demonstrated similar average completion times (7.8 and 7.6 minutes respectively), while Codex required significantly more time with an average of 16.6 minutes per generation task. The maximum observed generation time for Codex reached 1.74 hours for a single microservice, indicating potential timeout concerns for production use.

The time efficiency patterns reveal minimal difference between incremental and clean state generation scenarios for Claude Code and Code Qwen, suggesting these agents efficiently navigate both contexts. Codex showed higher variability, particularly in incremental P2 configurations where additional context increased generation time.

\subsubsection{Cost Efficiency}

\begin{figure}[tb]
    \centering
    \includegraphics[width=0.9\linewidth]{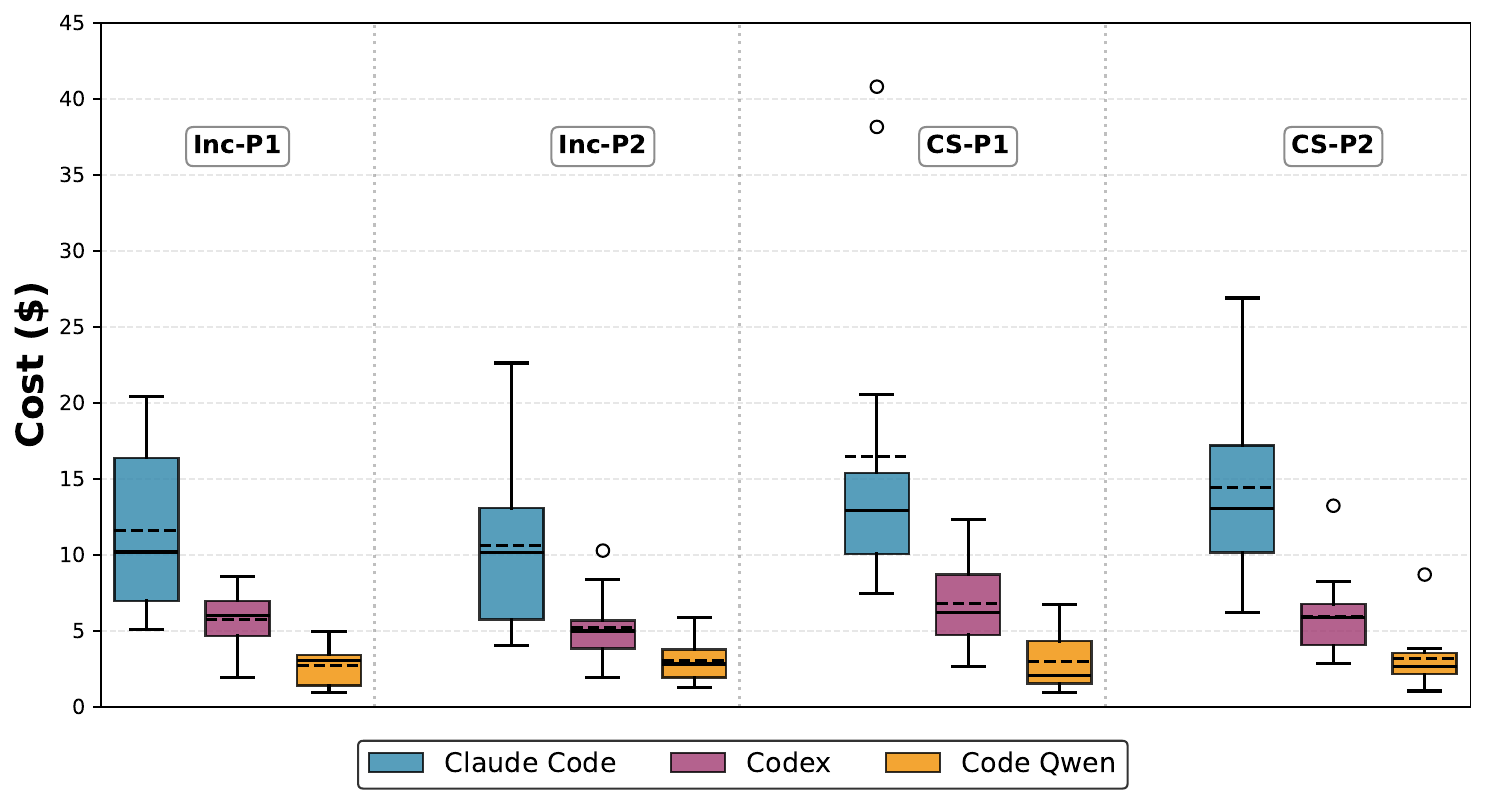}
    \caption{Cost distribution across agents and configurations}
    \label{fig:efficiency_cost}
\end{figure}

Figure~\ref{fig:efficiency_cost} presents the monetary cost distribution across configurations. Code Qwen emerged as the most cost-effective option with an average cost of \$2.98 per microservice generation, followed by Codex at \$5.92 and Claude Code at \$13.28. The total cost difference is substantial: generating all 48 microservices cost \$143 for Code Qwen, \$284 for Codex, and \$637 for Claude Code.

Despite higher per-token pricing, Claude Code's significantly lower output token count (averaging 2.1K tokens compared to 37.5K for Codex and 16.2K for Code Qwen) partially offset its higher input token costs. This suggests Claude Code exhibits more efficient token usage in its generation process.

\subsubsection{Token Consumption Patterns}

\begin{figure}[tb]
    \centering
    \includegraphics[width=\linewidth]{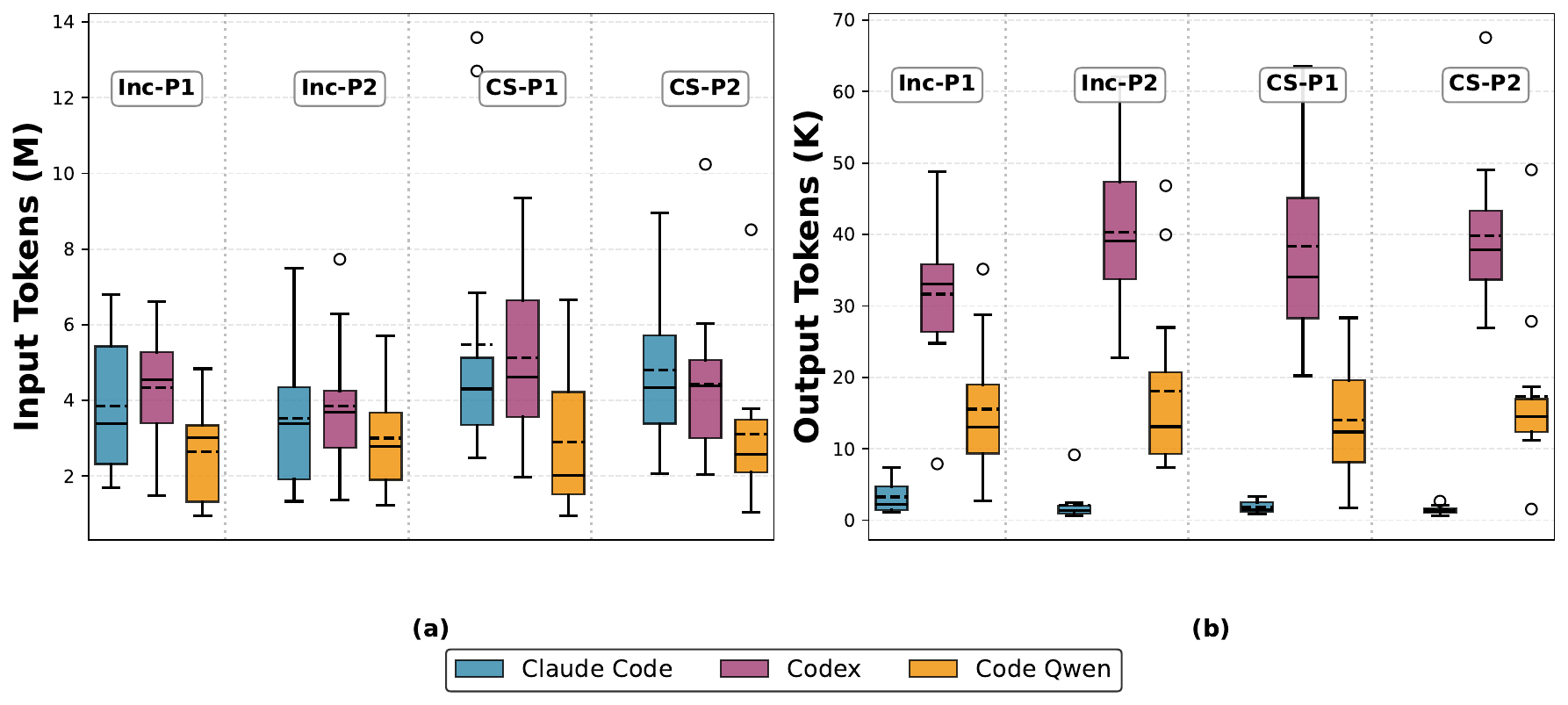}
    \caption{(a) Input and (b) Output Token distribution across agents and configurations}
    \label{fig:efficiency_tokens}
    \vspace{-7pt}
\end{figure}

Figure~\ref{fig:efficiency_tokens} show input and output token distributions respectively. Input token consumption was comparable between Claude Code (4.4M average) and Codex (4.4M average), while Code Qwen consumed 34\% fewer input tokens (2.9M average), likely due to its more efficient context processing.

% \begin{figure}[tb]
%     \centering
%     \begin{subfigure}[b]{0.48\linewidth}
%         \centering
%         \includegraphics[width=\textwidth]{images/efficiency_input_token_boxplot_merged.pdf}
%         \caption{Input tokens}
%         \label{fig:efficiency_input_tokens}
%     \end{subfigure}
%     \hfill
%     \begin{subfigure}[b]{0.48\linewidth}
%         \centering
%         \includegraphics[width=\textwidth]{images/efficiency_output_token_boxplot_merged.pdf}
%         \caption{Output tokens}
%         \label{fig:efficiency_output_tokens}
%     \end{subfigure}
%     \caption{Token distribution across agents and configurations}
%     \label{fig:efficiency_tokens}
% \end{figure}

Output token consumption showed dramatic differences: Claude Code generated remarkably concise outputs (2.1K tokens average), while Codex produced significantly more verbose code (37.5K tokens average). Code Qwen fell between these extremes at 16.2K tokens. When considered alongside the test pass rates from RQ$_{1}$, this suggests that verbose output does not necessarily correlate with higher functional correctness - Claude Code achieved comparable or better test pass rates despite generating substantially fewer output tokens.

\subsubsection{Efficiency Comparison}

\begin{figure}[t]
    \centering
    \includegraphics[width=\linewidth]{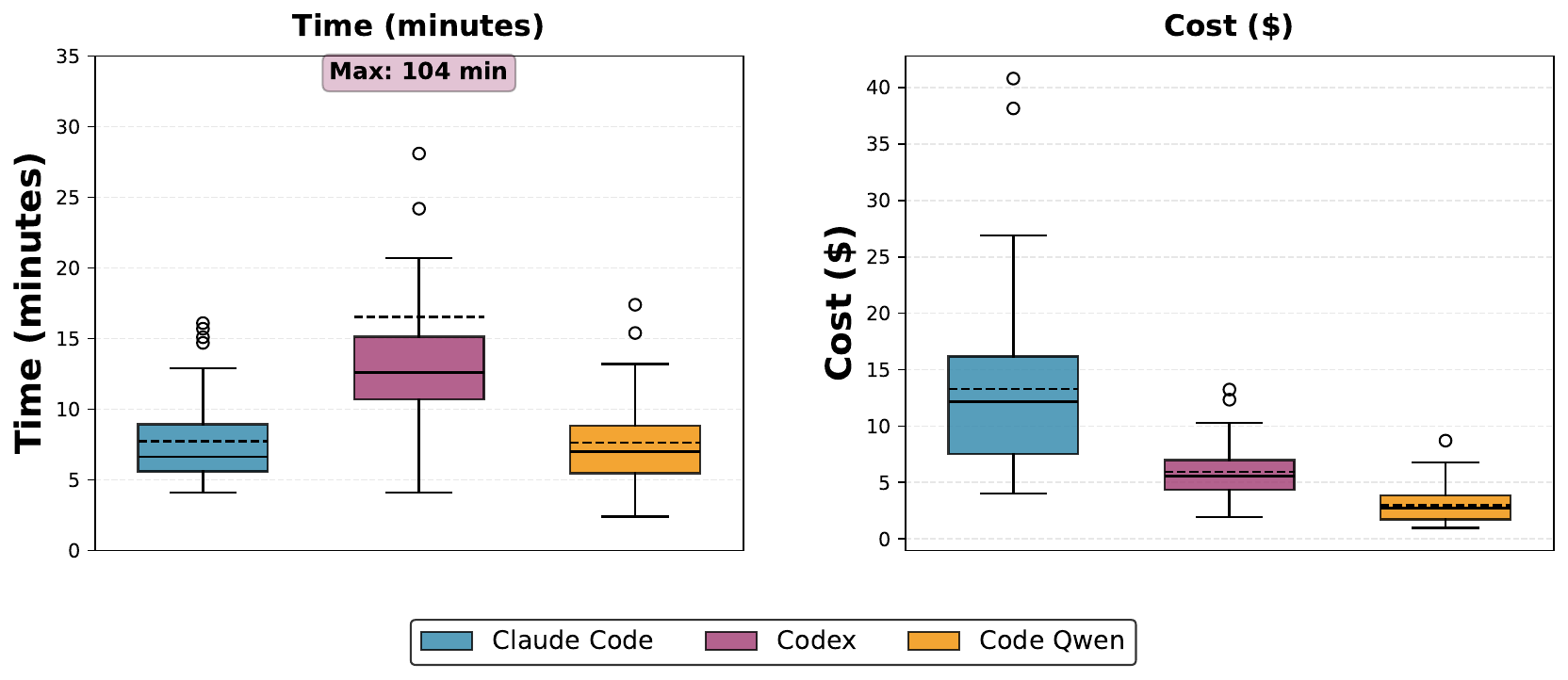}
    \caption{Time and Cost comparison across agents.}
    \label{fig:efficiency_compact}
    \vspace{-15pt}
\end{figure}

Figure~\ref{fig:efficiency_compact} provides an integrated overview of the primary efficiency dimensions. The time subplot (left) uses a truncated y-axis to highlight how Codex's extreme outliers (reaching 104 minutes) contrast sharply with the 0-20 minute range where most generations occur, while the cost subplot (right) illustrates the substantial cost differences across agents, with Code Qwen being most cost-effective and Claude Code being most expensive.

\begin{tcolorbox}[colback=orange!5!white, colframe=orange!95!white, colbacktitle=orange!95!white]
\small
    \textbf{Main Findings for RQ$_{2}$}: AI agents show significant differences in efficiency metrics. Claude Code and Code Qwen are faster (7-8 min average) while Codex is slower (16.6 min average) with concerning outliers reaching 1.74 hours. Claude Code is most expensive (\$13.28/service) while Code Qwen is most cost-effective (\$2.98/service) and generates the most concise code. Token verbosity does not correlate with functional correctness.
\end{tcolorbox}

 % Code Qwen is most cost-effective (\$2.98/service), while Claude Code is most expensive (\$13.28/service) but generates the most concise code. Token verbosity does not correlate with functional correctness.

\section{Discussion}\label{sec:discussion}
\subsection{Lessons Learned}

\textbf{Explicit implementation summaries degraded incremental generation.} Detailed summaries (P2) led to lower unit test pass rates in incremental scenarios compared to minimal prompts (P1), with Claude Code declining from 73.7\% to 63.2\% and Codex from 75.9\% to 50.3\%. Figure~\ref{fig:efficiency_tokens} reveals P2 prompts consumed fewer input tokens, indicating agents anchored on summaries rather than exploring the repository. Figure~\ref{fig:prompt-anchoring} illustrates this anchoring effect: for PiggyMetrics notification-service, Codex achieved 100\% test pass rate with P1 but degraded to 33.3\% with P2 by omitting MongoDB custom converters. While the summary mentioned "custom MongoDB converters for Frequency enum persistence" in implementation notes, it emphasized high-level concerns like OAuth2 integration, scheduled tasks, and service communication patterns. The agent anchored on these prominent features and generated a simplified configuration missing the low-level details it correctly inferred through exploration in P1. However, in clean state scenarios evaluated through integration tests, Code Qwen benefited substantially from explicit guidance. Integration tests only verify API contract compliance rather than internal implementation details, making them more tolerant of structural differences introduced when agents rely on summaries. These agent-specific and scenario-dependent sensitivities indicate that tromping strategies should be empirically determined rather than applying uniform approaches.

\begin{figure}[tb]
\centering
{\small \textbf{P1 Configuration (100\% test pass):}}
\begin{lstlisting}
@Configuration
static class CustomConversionsConfig {
    @Bean
    public CustomConversions customConversions() {
        return new CustomConversions(Arrays.asList(
            new FrequencyReaderConverter(),
            new FrequencyWriterConverter()));
    }
}
\end{lstlisting}

% \vspace{0.2cm}
{\small \textbf{P2 Configuration (33.3\% test pass - missing converters):}}
\begin{lstlisting}
@Bean
public Clock systemClock() {
    return Clock.systemUTC();
}
// CustomConversions configuration MISSING
\end{lstlisting}
\caption{Prompt-induced configuration loss in PiggyMetrics notification-service. With minimal prompt (P1), Codex correctly inferred MongoDB converters through codebase exploration (100\% test pass). With architectural summary (P2), the agent anchored on high-level patterns (OAuth2, scheduled tasks, service communication) mentioned prominently in the summary while omitting "custom MongoDB converters" mentioned only briefly in implementation notes, causing 12/18 tests to fail with \texttt{IllegalStateException}.}
\label{fig:prompt-anchoring}
\end{figure}
% \vspace{-10pt}
\noindent\textbf{Generated microservice implementation is typically more concise than baselines, but efficiency characteristics vary.} The microservices generated by agents were generally shorter than human-written baselines (see Figure~\ref{fig:code_quality_clean}), yet achieved comparable test pass rates. Generation times varied between the agents, with some showing concerning outliers. Agent costs ranged from \$2.98 to \$13.28 per microservice. The \$10 cost difference between the cheapest and most expensive agent is small enough that practitioners should prioritize correctness and reliability over cost efficiency when selecting agents.

\textbf{Data contamination effects were consistent across all agents.} Agents performed substantially better on popular open-source repositories (PiggyMetrics, Train-Ticket) than private projects, likely reflecting training data memorization or contamination. This disparity persisted across proprietary and open-source agents with different architectures, indicating systemic contamination. Our inclusion of student projects suggest that agents may underperform on proprietary codebases relative to benchmark expectations. For researchers, this underscores the necessity of private datasets for unbiased evaluation.
\vspace{-4pt}
\subsection{Implications for Practice}

\textbf{Agent and prompt selection must be determined empirically rather than following general guidelines.} Our results demonstrate no universally optimal configuration across scenarios. In incremental generation, minimal prompts (P1) consistently outperformed detailed summaries for Codex and Claude Code, while Code Qwen required explicit guidance (P2) in clean state scenarios to produce buildable code. Practitioners should conduct structured pilot evaluations: (1) select 2-3 microservices representative of their codebase complexity, (2) test both P1 and P2 prompts with their chosen agent, (3) measure test pass rates and generation time, and (4) establish baseline performance expectations before production deployment~\cite{shin2025prompt}. Agent-specific behaviors mean that general prompt engineering advice may not apply, and what works for one agent may degrade performance for another.

\textbf{Plan for timeout risks and establish generation time limits.} Codex showed concerning outliers reaching 104 minutes while the other agents took 7-8 minutes for generation on average. Production systems  should implement timeout mechanisms (e.g., 15-20 minute limits) and fallback strategies when generation exceeds acceptable thresholds or agents fail to produce output. For time-sensitive development workflows, select agents with the most predictable performance characteristics. Given that even the most expensive generation costs represent less than 15 minutes of developer time, reliability and correctness should be prioritized over minimizing costs.

\textbf{Expect performance degradation on proprietary codebases compared to benchmark results.} Organizations evaluating agents based on published benchmark results should anticipate lower performance on proprietary codebases, particularly for domain-specific business logic or uncommon architectural patterns. Pilot evaluations on actual production microservices provide more realistic performance expectations than public benchmarks. Additionally, generated code verbosity does not correlate with quality (Fig~\ref{fig:code_quality_clean}). Concise implementations can achieve comparable correctness to verbose implementations, so practitioners should not interpret brevity as insufficient or verbosity as thorough.

\noindent \textbf{Anticipate evolution toward architecture-specific tooling.} Current agents excel at implementation but lack architectural reasoning about organizational context and system-wide trade-offs. Emerging work by Jahića et al. ~\cite{jahic2025mindskillsgap} suggests the role of architects is shifting toward evaluating agent-generated solutions. Organizations should develop evaluation validation processes, as general-purpose agents may lack specialized architectural knowledge for complex system design decisions.

\vspace{-3pt}
\subsection{Implications for Research}

\textbf{Architecture-specific benchmarks are urgently needed to evaluate agent capabilities.} Existing benchmarks like SWE-bench \cite{swebench} focus on repository-level tasks but lack architectural components with explicit dependencies and API contracts. During our study, we faced challenges finding microservice systems with sufficient documentation and test suites. Future benchmarks should include: (i) diverse architectural styles beyond REST-based microservices, (ii) explicit API contract specifications, (iii) separate unit and integration tests, and (iv) varying complexity levels. Investigating few-shot learning with architectural examples~\cite{sun2025maad} and hybrid context strategies~\cite{longcodezip2025,repoformer2024} could give insights while designing the prompt.

\textbf{Methodologies for detecting and mitigating data contamination must be standardized.} Agents consistently performed better on popular repositories than private projects across, suggesting memorization rather than reasoning \cite{contamination_survey}. Since most LLM providers do not disclose training data, future research should develop contamination detection techniques through behavioral analysis.

\textbf{Human-agent collaboration frameworks require investigation to determine optimal intervention points.} Our results show varying success rates depending on context and complexity, yet when should humans intervene remains unclear. Research is needed to identify failure patterns through intermediate artifact analysis, establish risk-based review checkpoints, and develop iterative workflows where humans and agents refine implementations collaboratively. Trajectory analysis~\cite{bouzenia2025understandingsoftwareengineeringagents} provides foundation for understanding agent reasoning, but applying this to identify intervention triggers in architectural generation remains unexplored.

\noindent\textbf{Architecture-specific agents and specialized tooling warrant investigation.} Current agents are general-purpose code generators lacking architectural reasoning for system-level concerns like quality attributes and API contracts. Future research should explore specialized agents with architectural tools for contract validation, dependency analysis, and pattern enforcement to better support the complexity of architectural tasks.
\vspace{-3pt}
\subsection{Threats to Validity}

We follow the categorization provided by Wohlin et al. \cite{wohlin2012experimentation} and also provide brief explanation about efforts taken to mitigate the identified threats or why it is not possible, as suggested by Verdecchia et al. \cite{verdecchia_threats}.

\textbf{External Validity:} Selection of AI agents and generation parameters poses a threat to external validity. However, we systematically selected diverse agents based on LLM provider rankings \cite{chiang2024chatbotarenaopenplatform} and architecture-specific capabilities. Generalizability to other architectural patterns beyond REST-based microservices remains limited. Data contamination poses another threat, as the underlying LLMs may have been trained on the open-source repositories used in our study~\cite{contamination_survey}. Since most LLMs are not transparent about their training data, our inclusion of the  private projects implemented with lower visibility provides evidence of this effect, as agents consistently performed worse on these repositories compared to popular open-source projects.

\textbf{Internal Validity:} 
The use of different test strategies for incremental (unit tests) versus clean state (integration tests) scenarios poses a threat, as these validate different aspects of correctness. This distinction was necessary because agents generating from minimal context often produced functionally correct code with different structural implementations that would fail unit tests designed for specific baseline implementations (Fig~\ref{fig:cleanstate-mismatch}). Further, the use of Claude Code for the pilot study and summary generation poses a threat as it might optimize the performance of it, however Claude Code did not consistently outperform other agents across all scenarios.

\textbf{Construct Validity:} Selection of specific microservices for removal introduces a threat to construct validity. However, we selected 3 microservices per project based on business logic importance. Efficiency metrics were measured under controlled experimental conditions with similar network settings, though variations in cloud service latency may limit generalizability to local development environments. We do not account for prompt caching mechanisms, which may overstate actual costs for agents with effective caching, but all providers we use did not disclose cache-read/write pricing.

% \input{sections/threats_to_validity}

% \vspace{-5pt}
\section{Conclusion and Future Work}\label{sec:conclusion}
\vspace{-4pt}

This study assessed whether AI agents can generate functional and maintainable microservices (RQ$_1$) and how efficiently they operate (RQ$_2$). Across 144 generations, agents demonstrated the ability to produce microservices with code quality comparable to or simpler than human baselines. However, performance varied sharply by context: agents can generate functionally correct microservices in \~65\% of the cases when expected to be integrated into existing systems (\textbf{RQ$_{1.1}$}). Conversely, agents could generate functionally correct microservices in \~95\% of the cases  from requirements alone (\textbf{RQ$_{1.2}$}), where structural deviations did not hinder integration correctness. These results show that evaluation methodology and context availability strongly shape perceived capability.

Efficiency analysis (\textbf{RQ$_2$}) revealed large disparities. Claude Code and Code Qwen offered faster and more consistent generation, whereas Codex showed significant variance and costly outliers, and verbosity did not correlate with correctness. Overall, AI agents can support microservice development, yet their reliability remains scenario-dependent, and human oversight is required to ensure API and architectural consistency. Future work should establish architecture-specific benchmarks and clearer human–agent collaboration models to improve robustness and practical adoption.

\vspace{-5pt}
\section{Data Availability}
\vspace{-3pt}
The complete experimental dataset, including all 144 generated microservice implementations and prompts is available at \url{https://doi.org/10.5281/zenodo.17863951}.

% \balance
\bibliographystyle{ieeetr}
\bibliography{references}

\end{document}